\documentclass[11pt,a4paper]{article}
\usepackage{graphicx}
\usepackage{latexsym}
\usepackage{amsmath}
\usepackage{amssymb}
\usepackage{longtable} %long table
\usepackage{lscape} % rotate the figure or others
\usepackage{cite} % For reference, [1,2,3,4]=>[1-4]
\begin{document}
% Title and Authors
\title{\Large\textbf{A Circumstantial Evidence for the Possible Production
of QGP in the 158A GeV/c Central Pb+Pb Collisions}
}%
\author{Zhiyi Liu, Benhao Sa, Shuhua Zhou\\ \small{China Institute of Atomic Energy, P.O.Box 275(18), 102413}
\\ \underline{zyliu@iris.ciae.ac.cn}}%
\maketitle
%\date{Otc 18, 2003}%
% Abstract
\begin{abstract}
Hadron and string cascade model (JPCIAE) with the hypothesis without introducing the quark-gluon plasma (QGP), is employed to study the direct photon and $\pi^0$ transverse momentum distributions for central $^{208}$Pb+$^{208}$Pb collisions at 158A GeV/c . JPCIAE model, is based on LUND model, especially on the envent generator PYTHIA, and can be used to simulate the relativistic nucleus-nucleus collisions where PYTHIA is called to deal with hadron-hadron collisions. In our work, the theoretical results of transverse momentum distribution for both the direct photon and the $\pi^0$ particle are lower than the data of WA98 experiment. However, JPCIAE model can ever explain successfully the results of WA80 and WA93 experiments of central S+Au collisions at 200A GeV/c where no evidence of direct photon excess. Having considered the results of WA80 and WA93 experiments can be explained but WA98's can't , that might provide a circumstantial evidence for the possible production of QGP in the high-energy central Pb+Pb collisions.\end{abstract}
% -------------------- Formal Document ------------------------
%\subsection{}
%\subsubsection{}
%\paragraph{}
%\subparagraph{}
One of the aims to research relativistic heavy-ion collisions is to explore the probability of quark-gluon plasma phase transition from normal nuclear matter. There have already accumulated some important results since the experiments of 15A GeV/c nuclear collisions in BNL-AGS begun at 1980's till the experiments of 200A GeV/c nuclear collisions in CERN-SPS begun at 1990's.

 There are proposed signatures to diagnose the production of QGP in relativistic nucleus-nucleus collisions. Among those the signatures related with final state are: the strangeness enhancement, the anomalous J/$\psi$ suppression and the excess of direct photon etc.\cite{00Won}. The collected data from AGS up to SPS energies have already leaded some of us to conclude that we have now specious evidence that a new state of matter has indeed created.

Earlier job about this is from Ref.\cite{96Bla,96Won}. They assume that the anomalous suppression of J/$\psi$ production in Pb+Pb collisions at 158 A GeV/c can be explained by introducing QGP formation only. At the beginning of the year 2000 a news released from CERN stated that a new state of matter, QGP, was created by high-energy nuclear collisions at CERN\cite{00Hei}.

In the quark-gluon plasma phase, particles including photons will be emited. Photons, also those arising from the electromagnetic interactions of the constituents of the plasma, will provide information on the properties of the plasma. Since the small reinteraction cross section of photon with surround medium, the photon is expected to be a relatively 'clean' probe to study the state of the QGP. The presence of these photons in high-energy heavy-ion collisions can also provide possibly evidence for the production of QGP. Photons are also produced by many other hadronic processes in heavy-ion reactions, such as the decay of  and  , they are much dominated indeed. Only those photons after the subtraction of the photons from hadronic decay are meaningful in diagnosis of the QGP, which refers to as 'direct photons'.

Ref.\cite{00Won} employed the perturbative QCD and Glauber model to analyze the photon excess in central Pb+Pb collisions at 158A GeV/c\cite{00Agg}. The good agreement between the model and the WA98 data reached only if one assumes the production of QGP in high-energy central Pb+Pb collisions.

High energy photon emission rate from matter created in Pb+Pb collisions at CERN SPS energies was also evaluated within the framework of (3+1) dimensional hydrodynamic expansion in Ref.\cite{00Alb}. The photon spectra measured by the WA98 experiment were well reproduced by hard QCD photons and photons from a thermal source, QGP, with initial temperature ~200 MeV.

WA80 and WA93 collaborations have measured direct photons in 200A GeV/c S+Au collisions prior to the WA98 experiment, but their results showed no significant direct photon excess\cite{96Wa80,97Wa93}.

Ref.\cite{00Wan} ever studied carefully the data of the WA80 and WA93 experiments using hadron and string cascade model (JPCIAE). Their results not only reproduce successfully the WA80 data of upper limits of the direct photon transverse momentum distribution but also explain the WA93 data of the inclusive photons low transverse momentum distribution. Hadron and string cascade model (JPCIAE) , presupposing the hypothesis without introducing the quark-gluon plasma (QGP) , is based on LUND model, on the envent generator PYTHIA especially, and is used to simulate the relativistic nucleus-nucleus collisions where PYTHIA is called to deal with hadron-hadron collisions. As a extension of Ref.\cite{00Wan}, we discuss, in this paper, the transverse momentum spectra of direct photons produced in 158A GeV/c Pb+Pb central collisions using JPCIAE model.

\begin{figure}
  % Requires \usepackage{graphicx}
  \centering
  \includegraphics[width=8cm]{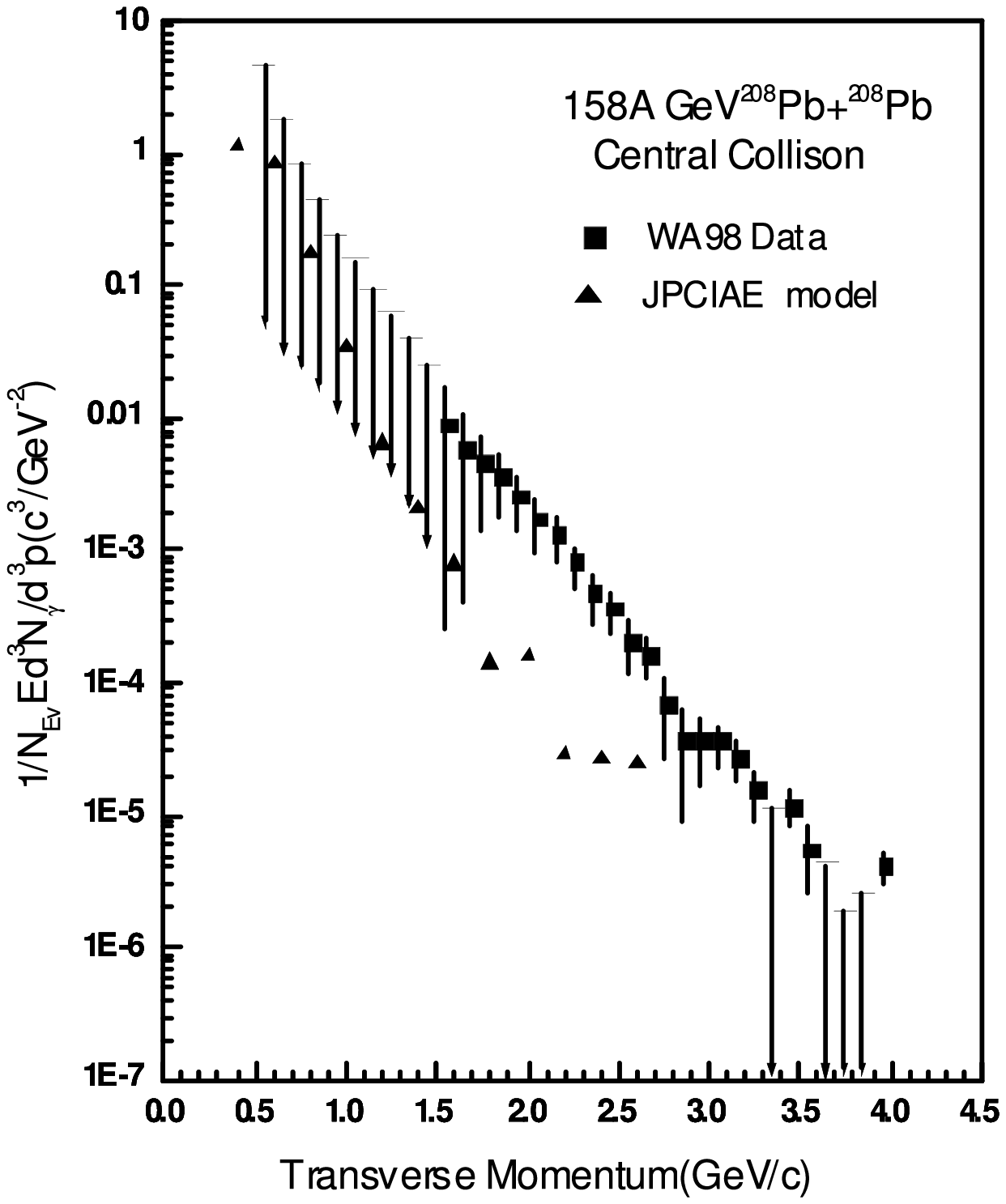}\\
  \caption{The transverse momentum distribution of direct photons
in Pb+Pb central collisions at 158A GeV/c
}\label{ph}
\end{figure}

In Fig.\ref{ph} the full squares with the error bar are the experimental transverse momentum distribution of direct photon in 158A GeV/c Pb+Pb central collisions, and the full triangles are the corresponding results of JPCIAE. As the same as other transport models (HIJING, RQMD and VENUS for instance), JPCIAE model is hard to calculate the transverse momentum spectra above Pt=3.0 GeV/c. However, from the trend of the distributions in region of Pt$<$3.0 GeV/c we can conclude that both the distribution of JPCIAE model and WA98 are the same globally in trend but theoretical results are a factor of 1-2 magnitude lower than the WA98 data.
\begin{figure}
  % Requires \usepackage{graphicx}
  \centering
  \includegraphics[width=8cm]{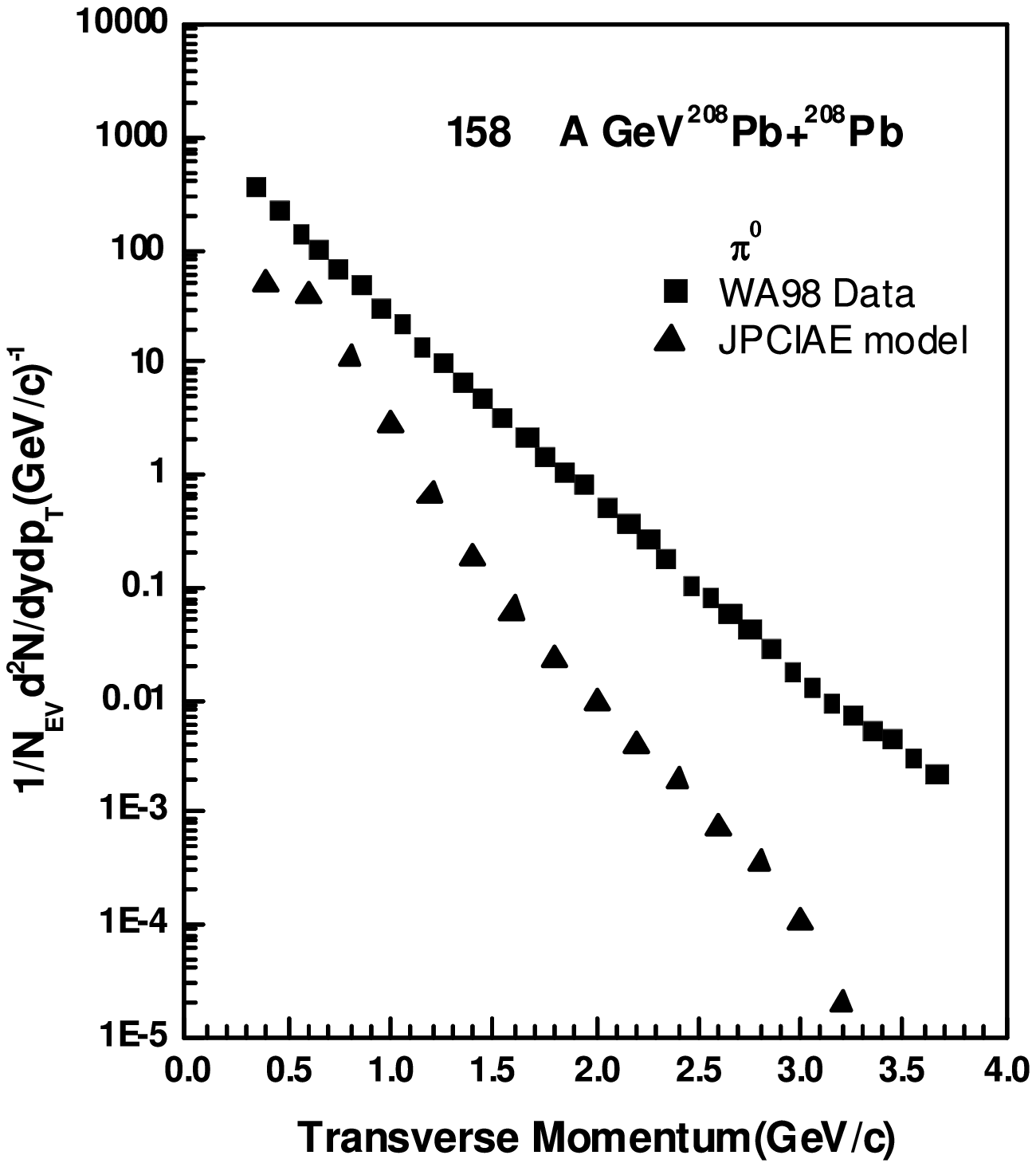}\\
  \caption{$\pi^0$ transverse momentum in Pb+Pb central collisions }\label{pi}
\end{figure}
The full squares, in Fig.\ref{pi}, are the transverse momentum distribution of $\pi^0$ in 158A GeV/c Pb+Pb central collisions and the full triangles are the corresponding results of JPCIAE. One sees from this figure that the theoretical results of $\pi^0$ transverse momentum distribution are also lower than WA98 data and the discrepancy between theory and experiment increases with the increasing of Pt. It indicates , to some extent, that the possibility of producing QGP at high Pt region is bigger.

As a whole, the theoretical results of transverse momentum distribution for both the direct photon and the $\pi^0$ particle are lower than the data of WA98 experiment. However, JPCIAE model, where the quark-gluon plasma (QGP) is not introduced, explicitly can ever explain successfully the results of WA80 and WA93 experiments of central S+Au collisions at 200A GeV/c where no evidence of direct photon excess. Having considered the results of WA80 and WA93 experiments can be explained but WA98's can't , that might provide a circumstantial evidence for the possible production of QGP in the high-energy central Pb+Pb collisions.

% -------------------------------------------------------------
% ----------------- Beginning of Bibliography -----------------

% -------------------------------------------------------------
\end{document}